\documentclass{aa}
\usepackage{epsfig}
\usepackage{rotating}
\usepackage{natbib}
\usepackage{graphicx}

\newcommand{\xmm}{{\sl XMM}}
\newcommand{\hst}{{\sl HST}}

\newcommand{\vlt}{{\sl VLT}}
\newcommand{\ntt}{{\sl NTT}}

\newcommand{\fors}{{\sl FORS}}
\newcommand{\forsone}{{\sl FORS1}}
\newcommand{\forstwo}{{\sl FORS2}}
\newcommand{\vltn}{{\sl Very Large Telescope}}

\def\gasm{\hbox{\hspace{0.5mm}}\raise2pt
       \vbox{\hbox{$>$}}\lower2pt
              \vbox{\moveleft6.0pt\hbox{$\sim$ }}\hbox{\hskip 0.05mm}}

\begin{document}

  \title{The Optical Spectrum of the Vela Pulsar\thanks{Based on observations collected at the European Southern Observatory, Paranal, Chile under  programme ID 66.D-0261(A) } }

 \author{R.P. Mignani\inst{1}
 \and 
	S. Zharikov\inst{2}
\and 
	P. A. Caraveo\inst{3}
}

   \offprints{R.P. Mignani}

   \institute{Mullard Space Science Laboratory, University College London, Holmbury St. Mary, Dorking - Surrey, RH5 6NT, UK\\
              \email{rm2@mssl.ucl.ac.uk}
\and
Instituto de Astronomía, Universidad Nacional Aut\'onoma de M\'exico, Apartado Postal 877, 22830, Ensenada, Baja California, M\'exico \email{zhar@astrosen.unam.mx}
\and
INAF, Istitituto di Astrofisica Spaziale, Via Bassini 15, Milan, 20133 Italy \\
\email{pat@iasf-milano.inaf.it} }

     \date{Received ...; accepted ...}

   \abstract{Our knowledge of the  optical spectra of Isolated Neutron
Stars (INSs)  is limited by  their intrinsic faintness.  Among  the fourteen
optically  identified  INSs,   medium  resolution  spectra  have  been
obtained  only  for  a  handful  of objects.   No  spectrum  has  been
published yet for the Vela pulsar (PSR B0833--45), the third brightest
($V=23.6$)  INS  with  an  optical  counterpart.   Optical  multi-band
photometry underlines a flat continuum.
}{In  this work we present  the first optical
spectroscopy  observations  of  the  Vela  pulsar,  performed  in  the
4000-11000\AA\ spectral  range.}{Our observations have  been performed
at  the  ESO \vltn\  (\vlt)  using  the  \forstwo\ instrument.   }{The
spectrum  of the  Vela pulsar  is  characterized by  a flat  power-law
$F_{\nu}  \propto  \nu^{-\alpha}$  with  $\alpha  =  -0.04  \pm0.04  $
(4000-8000  \AA), which  compares  well with  the values obtained  from
broad-band  photometry.  This  confirms, once  more, that  the optical
emission of  Vela is entirely  of magnetospheric origin. 
}
{The   comparison   between    the   optical   spectral   indeces   of
rotation-powered INSs does not  show evidence for a spectral evolution
suggesting that, as  in the X-rays, the INS aging  does not affect the
spectral properties of the  magnetospheric emission. At the same time,
the optical  spectral indeces  are found to  be nearly  always flatter
then the  X-rays ones, clearly suggesting a  general spectral turnover
at lower energies.}

   \maketitle

\section{Introduction} 
The study of the optical emission properties of Isolated Neutron Stars
(INSs) is hampered by their  intrinsic faintness. For most of them the
knowledge of  the optical  spectrum is still  based on  the comparison
between multi-band  photometry measurements.   We note here  that this
comparison  is  affected  by  several uncertainties  since  photometry
measurements  are often compiled  from the  literature and,  thus, are
taken  with   different  instruments  and   filters,  calibrated  with
different photometric systems  (167 used in total; see,  e.g.  Mono \&
Munari           2000\footnote{http://ulisse.pd.astro.it/Astro/ADPS/}),
indipendently   corrected  for   the   atmospheric  and   interstellar
extinction and, last but not least, converted to spectral fluxes using
slightly  different techniques.  Only  for five  of the  fourteen INSs
with  identified  optical  counterparts  (see Mignani  et  al.   2004;
Mignani 2005 for updated  reviews) medium-resolution spectra have been
obtained so  far.  The first one  was the Crab pulsar  (Oke 1969), the
youngest  ($\sim 1000$ years)  and the  brightest INS  ($V=16.6$), for
which repeated spectroscopy  observations have been performed (Nasuti
et  al.   1996; Sollerman  et  al.  2000).   The  second  one was  PSR
B0540--69 (Hill et al.  1997, Serafimovich et al.  2004), the youngest
($\sim  2000$  years) and  brightest  ($V=22.4$)  INS  after the  Crab
pulsar.  Only  after the advent  of the generation  of the 10  m class
telescopes, it was  possible to obtain spectra of  the fainter ($V\sim
25-25.6$) and older ($\ge 10^5$ years) INSs such as Geminga (Martin et
al.  1998) and PSR B0656+14 (Zharikov et al.  2007), as well as of the
radio-quiet INS RX J1856--3754 (van Kerkwijk \& Kulkarni 2001).

The best next target for  optical spectroscopy is the Vela Pulsar (PSR
B0833--45), the  third optically brightest ($V=23.6$)  INS.  Among the
first   radio  pulsars   discovered  in   the  sixties,   its  optical
identification  was proposed  by Lasker  (1976) and  confirmed  by the
detection  of  optical pulsations  at  the  radio  period (Wallace  et
al. 1977).   The high pulsed  fraction of the optical  lightcurve (see
also  Gouiffes  1998),  naturally   pointed  towards  a  non  thermal,
magnetospheric origin of the optical radiation.  This was confirmed by
multi-band  photometry performed  by Nasuti  et al.   (1997)  with the
\ntt\  and by Mignani  \& Caraveo  (2001) with  the \hst.   The fluxes
between  $\sim3000$  \AA\  and  $\sim8000$  \AA\  follow  a  flattish,
power-law distribution ($F_{\nu}  \propto \nu^{-\alpha}; \alpha = -0.2
\pm 0.2$).   The non-thermal  nature of the  optical radiation  of the
Vela  pulsar  has been  also  confirmed  by polarisation  observations
performed  with the  \vlt\ (Wagner  \&  Seifert 2000;  Mignani et  al.
2007),  pointing to a  relatively high  polarisation level.   The Vela
pulsar has been  recently observed in the IR  with the \vlt\ (Shibanov
et al.  2003), and in the  near-UV (Romani et al.  2005; Kargaltsev \&
Pavlov  2007) with  the  \hst.  While  its  multi-band photometry  now
extends  from the  IR to  the near-UV,  no optical  spectrum  has been
published yet.  Here, we report on the first spectroscopy observations
of  the  Vela pulsar  performed  with  the  \vlt.  Observations,  data
analysis and  results are described in \S2,  while the interpretations
are discussed in \S3.

\section{Observations and Data Analysis}

\subsection{Observations}

Spectroscopy observations of the Vela pulsar were performed in Service
Mode in four  different nights between December 2000  and January 2001
with the \vlt\ at the ESO's Paranal Observatory (Chile).  We have used
\forstwo\                        (FOcal                        Reducer
Spectrograph\footnote{www.eso.org/instruments/fors}),   a   multi-mode
camera for imaging and  long-slit/multi-object spectroscopy which is 
identical to \forsone\ (Appenzeller  et al. 1998) but with the CCD
optmized  to achieve  a higher  sensitivity in  the Red  part  of the
spectrum.   At  the  time  of  the observations  \forstwo\  was  still
mounting  the original detector,  a Tektronix  2048$\times$2046 pixels
CCD with  a plate  scale of $0\farcs20$  with the  Standard Resolution
(SR)  collimator.  The
instrument was operated in its single-port, high-gain, read out mode which
is the  default one  for long slit  spectroscopy (LSS).  To  cover the
wavelength  interval 4000-11000\AA\  the  observations were  performed
with two different grisms: the $300V$ ($\Delta \lambda=4500-8600$ \AA)
and the  $300I$ one  ($\Delta \lambda=6000-11100$ \AA),  equipped with
the order separation filter $OG590$.  Both grisms have a dispersion of
$\sim$2.7\AA/pixel and  a resolving power $\lambda /  \Delta \lambda =
440$.  A slit width of $2\farcs5$  was used to collect as much flux as
possible from the pulsar (see Fig. 1).

\begin{figure}[h]
\centering           
\includegraphics[bb=50 150 550 650,width=8.0cm,clip]{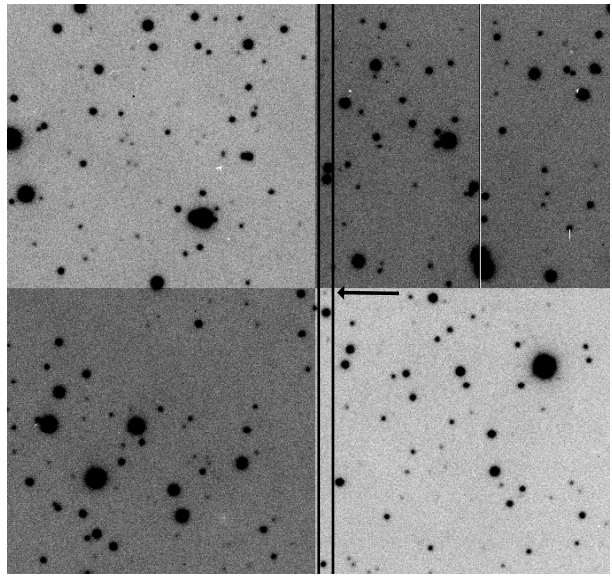}
\caption{VLT  \forstwo\ raw  $B$-band  acquisition image  of the  Vela
pulsar.  North to the top, East  to the left.  The $2\farcs5$ LSS slit
is overplotted,  with the  orientation NS as  used in the  first night
(see Table  1).  The position of  the Vela pulsar  (barely detected in
the  60 s exposure  acquisition image)  is marked  by the  arrow.  The
difference between the four  quadrants corresponds to the different CCD
readout ports. }
\label{vela_acq}       
\end{figure}

\begin{figure*}
\centering           
\includegraphics[bb=-150 300 763 500,width=14.0cm,clip]{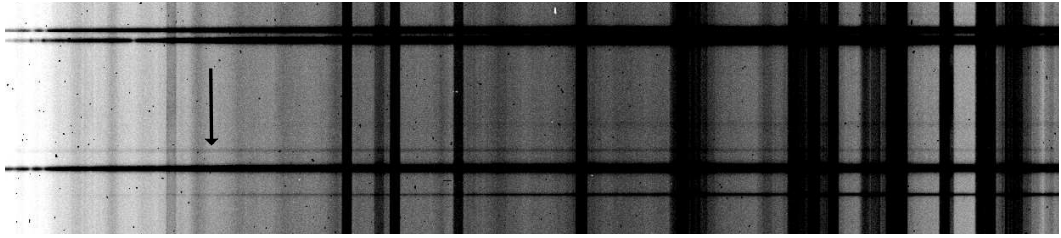}
\caption{VLT/\forstwo\  two-dimensional spectrum  (first  night) taken
through  the $300V$  grism and  the $2\farcs5$  slit oriented  NS (see
Figure 1). The Vela pulsar spectrum is marked by the arrow.  }
\label{vela_2dspec}       
\end{figure*}

Single LSS  science exposures  of 2800 s  each were obtained  for both
grisms  and repeated  for cosmic  rays  filtering.  A  total of  eight
2800~s exposures  were taken for  a total integration time  of 11200~s
for  each grism.   The complete  log  of the  science observations  is
reported in Table  1. The first exposure sequence  in the $300I$ grism
was  aborted and  is not  considered in  the following  analysis.  The
seeing conditions were always  sub-arcsec ($0\farcs8$ average) and the
airmass below  1.2.  The slit was  oriented NS on the  first night and
always  EW   on  the  following  nights.   For   each  night,  daytime
calibrations (biases, darks, flatfields,  arc lamp spectra) were taken
to  correct for  instrumental effects  and to  perform  the wavelength
calibration.   Multi  object spectroscopy  (MOS)  observations of  the
spectrophotometric  standard stars  LTT 3218,  Feige 56  and  Feige 67
(Hamuy et al. 1994) were acquired at the beginning of each night (with
the only exception of the third one) for flux calibration.  Because of
their brightness,  standard stars were  observed in MOS  mode (slitlet
size $22\farcs0$)  to avoid flux  losses which might occur  when using
the narrower  LSS slit ($2\farcs5$).  Thanks to  the sub-arcsec seeing
conditions  during the  observations it  was possible  to  resolve the
pulsar spectrum  in each individual science exposure.   As an example,
we show  in Fig. 2 one  of the two  $300V$ spectra taken on  the first
night,  where  the pulsar  is  clearly  detected.  Unfortunately,  the
pulsar  is detected  with  a  much lower  significance  in the  $300I$
spectra due  to the higher sky background  and to the drop  of the CCD
sensitivity towards longer wavelengths

\begin{table}
\begin{center}
\caption{Summary of  the spectroscopy observations of  the Vela pulsar
taken with \forstwo. The columns give the observing dates, the used grism, the exposure times, the number of exposures per pointing N, and the average seeing and airmass values during each exposure sequence. }
\begin{tabular}{cccccc} \hline
yyyy-mm-dd & Grism & Time (s) & N  & Seeing  & Airmass	\\ \hline
2000-12-03 & 300V   & 2800 & 2 &$0\farcs79$ & 1.14 \\ 
2001-01-23 & 300V   & 2800 & 2 &$0\farcs60$ & 1.07\\  
2001-02-25 & 300I   & 1400 & 1 &$0\farcs89$ & 1.09\\  
2001-01-30 & 300I   & 2800 & 4 &$0\farcs73$ & 1.11\\ \hline
\end{tabular}
\label{tabdatasummary}
\end{center}
\end{table}

\subsection{Data Analysis}

Spectral data reduction and  calibration were performed using standard
suites of tools for treating CCD  data available in the MIDAS and IRAF
sofware packages.  Master calibration frames (master biases and darks,
flux-normalized  flat  fields)  were   provided  by  the  \fors\  data
reduction
pipeline.\footnote{http://www.eso.org/observing/dfo/quality/FORS1}
Science and standard stars spectra,  as well as arc lamp spectra, were
then   bias   and   dark-subtracted,  and   flatfielded.    Wavelength
calibration was computed from the  reduced arc lamp spectra by fitting
a  second order  polynomial, yielding  an rms  of 0.4  \AA/pixel.  The
wavelength  calibration  was  then  applied  row-by-row  to  both  the
standard  stars  and  science  spectra.   For  each  grism,  the  flux
normalization was computed from the extracted one-dimensional standard
star spectrum  using the available  flux reference table  and applying
the  atmospheric  extinction  correction  with the  extinction  curves
measured for  the Paranal Observatory  (Patat 2004).  For  each grism,
single science spectra were  finally co-added and cosmic rays removed.
The pulsar  one-dimensional spectrum was  then extracted from  each of
the  co-added  science  spectra  using  a 4  pixel  wide  ($0\farcs8$)
extraction window  centered on  the pulsar position  where the  S/N is
higher.  The  sky background  was computed and  subtracted from  two 6
pixel  wide ($1\farcs2$)  regions immediately  adjacent to  the pulsar
spectrum.  For  each extracted one-dimensional spectrum  of the pulsar
we  finally  applied  a  flux  correction factor,  computed  from  the
extracted spectra  of brighter field  stars, to account for  the width
difference between  the extraction window  and the object's  PSF.  The
spectra  were corrected  for the  interstellar extinction  using  as a
reference $E(B-V)=0.05$  obtained from the  spectral fit to  the \xmm\
X-ray spectrum  (e.g. Manzali et al.  2007).   The two one-dimensional
spectra of the  pulsar were finally merged to  yield a single spectrum
across the 4000-11000\AA\ wavelength range.

\begin{figure*}
\centering           
\includegraphics[bb=120 15 544 770,width=7.0cm,angle=-90,clip]{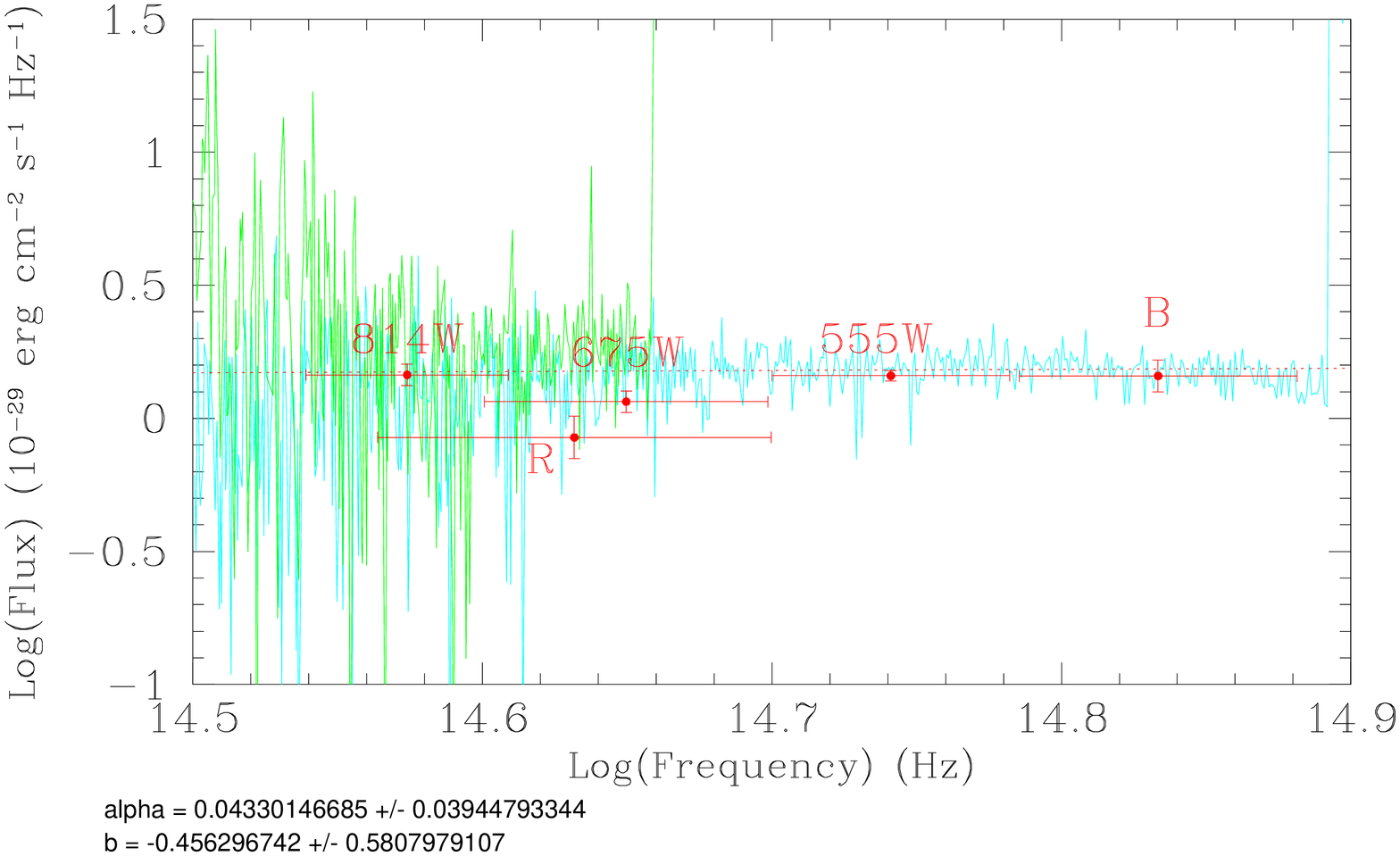}
\caption{Extracted  one-dimensional   spectrum  of  the   Vela  pulsar
obtained  by  merging the  $300V$  and   $300I$   spectra (shown in green and cyan, respectively)  after
wavelength/flux  calibration, extinction correction,  and  coaddition. A rebinning of a factor 3 has been applied to both spectra. The spikes
are due to defects in the subtraction of bright sky lines and of residual cosmic ray hits. The red points correspond to broad-band photometry (Nasuti et al. 1997; Mignani  \& Caraveo 2001).  The red dotted
line represents the best fitting power-law ($\alpha = -0.04 \pm 0.04$) to the continuum.
Since  at longer  wavelengths the  spectrum is  dominated by  the sky
background,  the  fit  has  been  performed only  between 4000 and 8000 \AA. }
\label{vela_2dspec}       
\end{figure*}

\subsection{Results}

The final  Vela pulsar's  spectrum is  shown in Fig.   3.  Due  to the
different detection significance of the  pulsar in the two grisms, the
S/N of the resulting  (unbinned) spectrum is rather unhomogeneous.  In
particular,  we  estimated   a  S/N  $\sim$  5  (per   pixel)  in  the
4000-5500\AA\ wavelength range and a S/N as low as $\sim 1$ near/above
8000\AA.  At longer wavelengths  the spectrum is entirely dominated by
noise. For  this reason in  the following analyses we  have considered
only  the  part  of the  spectrum  between  4000  \AA\ and  8000  \AA.
Clearly,  the spectrum  consists of  a  pure flat  continuum, with  no
evidence  for emission  or absorption  feautures.  For  comparison, we
have  overplotted  the spectral  fluxes  recomputed  from the  optical
multi-band photometry  (Nasuti et al. 1997; Mignani  \& Caraveo 2001),
obtained  between  $\sim$  4000  \AA\  and $\sim$  8000  \AA.   To  be
consistent, we  have re-corrected these  fluxes for the same  value of
the interstellar extinction used in this work ($A_V$=0.16), while both
Nasuti  et  al.   (1997)  and  Mignani \&  Caraveo  (2001)  have  used
$A_V$=0.4. For the interstellar extinction correction we have used the
coefficients   of   Fitzpatrick   (1999).    As   seen,   within   the
cross-calibration  uncertainties, the  multi-band  photometry and  the
spectroscopy flux measurements  are substantially consistent with each
other.   A putative  dip at  6500 \AA\  was hinted  in  the multi-band
photometry data  of Mignani \& Caraveo (2001).  However, its existence
was   not  confirmed   by  the   \vlt\  photometry   of   Shibanov  et
al. (2003). No  evidence for such a dip is found  in our spectral data
either.   We  thus  conclude  that   it  was  just  the  result  of  a
cross-calibration problem  in the multi-band photometry  of Mignani \&
Caraveo (2001).

We have fitted the \forstwo\  spectrum between 4000 and 8000 \AA\ with
a  power-law  ($F_{\nu}  \propto  \nu^{-\alpha}$)  and  we  derived  a
spectral index  $\alpha =-0.04 \pm  0.04$.  This value  superseeds the
one  obtained from  the multi-band  photometry of  Mignani  \& Caraveo
(2001) on the base  of only five spectral flux  measurements.  We note
that our spectral  index is somewhat flatter than  the one obtained by
Shibanov et  al.  (2003) by fitting the  IR-to-optical spectral fluxes
($\alpha = 0.12 \pm 0.05$), while it compares better with the spectral
index  $\alpha =  0.01 \pm  0.02$,  obtained by  Kargaltsev \&  Pavlov
(2007) by extending the fit to the near-UV.
Interestingly, the optical power-law is below the extrapolation of the
power-law component  ($\alpha =  1.2 \pm 0.3$)  used to fit  the \xmm\
X-ray spectrum  (Manzali et al.   2007; see also Kargaltsev  \& Pavlov
2007), which clearly suggests  that the magnetospheric emission of the
Vela pulsar features a spectral turnover at longer wavelengths.

\begin{figure}
\centering   
\includegraphics[bb= 1 -30 593 838, width=8.0cm,clip=]{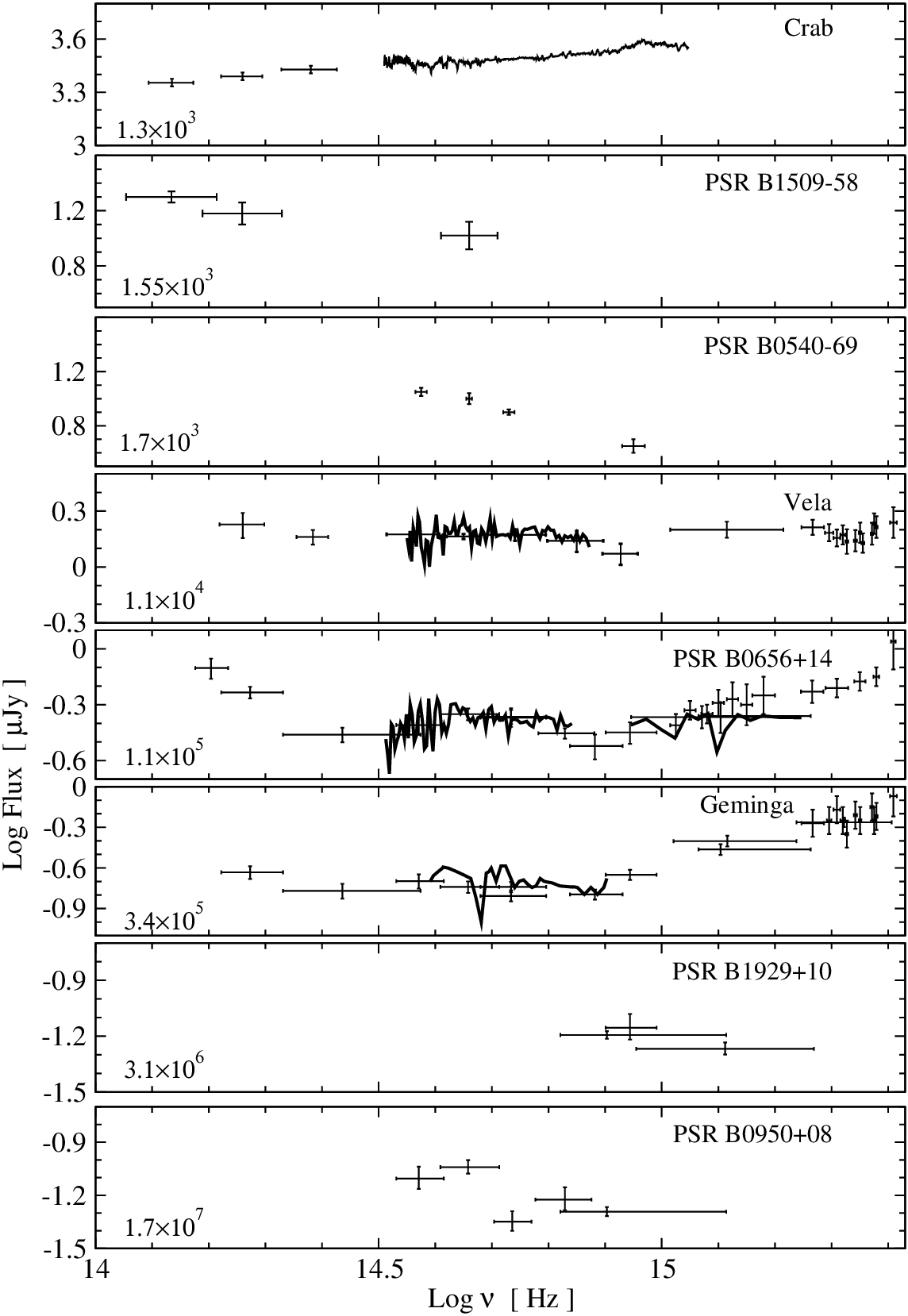}
\caption{Spectral  flux distribution  of all  rotation-powered pulsars
for   which  either   medium-resolution  spectroscopy   or  multi-band
photometry is available (see Table 2). From top to bottom, objects are
sorted  according to  increasing  spin-down age.  This  figure has  been
updated from  Fig. 1 of  Zharikov et al.  (2007), by including  the IR
fluxes of PSR B1509--58 (Kaplan \& Moon 2006), the Vela spectrum (this
work), and  the near-UV fluxes of  Vela (Romani et  al. 2005), Geminga
(Kargaltsev  et al.  2005) and  PSR  B0656+14 (Shibanov  et al.  2005;
Kargaltsev \& Pavlov 2007).}
\label{multispec}       
\end{figure}

\section{Discussion}

\begin{table*}
\begin{center}
\caption{Summary of the measured optical/IR spectral index $\alpha_O$ of rotation-powered INSs as obtained from spectroscopy (column three) and multi-band photometry (column five).  Columns 2-4 give (in logharitmic units) the pulsar spin down age $\tau$, the magnetic field $B$, and the rotational energy loss $\dot E$. The last column gives the phase-averaged X-ray spectral index $\alpha_X$.}
\begin{tabular}{lccc|cc|cc|c|l} \hline
Name & Log($\tau$) & Log($B$) & Log($\dot E$) & $\alpha_{O,sp}$ & $\lambda\lambda$   & $\alpha_{O,ph}$ & $\lambda\lambda$  & $\alpha_X$ & Ref.\\ 
   & yrs & (G) & (erg s$^{-1}$)        & &(\AA)  & & (\AA)  &            & \\ \hline
Crab         &3.1 & 12.58 & 38.65 &$-0.11\pm0.04$ & 3300-9250  &               &            & $1.079\pm0.004$ &1,2 \\ 
B1509--58    &3.2 & 13.19 & 37.25 &               &            & $0.5$         & 6000-18000 & $0.4\pm0.5$     &3,4 \\
B0540--69    &3.2 & 12.70 & 38.17 &$1.88\pm0.01$  & 3605-6060  & $1.07\pm0.2$  & 3300-8000  & $0.88\pm0.11$   &5,4\\  
Vela         &4.1 & 12.53 & 36.84 &$-0.04\pm0.04$ & 4000-8000  & $0.01\pm0.02$ & 1500-16000 & $1.2\pm0.3$     &6,7,8 \\  
B0656+14     &5.0 & 12.67 & 34.58 &$0.2\pm0.2$    & 4600-7000  & 0.41          & 4300-18000 & $1.1\pm0.3$     &9,10,11 \\ 
Geminga      &5.5 & 12.21 & 34.51 &$0.8\pm0.5$    & 3700-8000  & 0.46          & 4300-16000 & $0.7\pm0.1$     &12,10,11\\ 
B1929+10     &6.5 & 11.71 & 33.59 &               &            & $0.5\pm0.5$   & 1700-3400  & $1.7\pm0.1$     &13,14 \\
B0950+08     &7.2 & 11.39 & 32.75 &               &            & $0.65\pm0.4$  & 3600-8000  & $0.9\pm0.1$     &15,16 \\\hline
\end{tabular}
\label{tabdatasummary}
\end{center}

(1) Sollerman et al. (2000), (2) Kirsch et al. (2006), (3) Kaplan \& Moon (2006),(4) Gotthelf (2003),(5) Serafimovich et al. (2004), (6) this work, (7) Kargaltsev \& Pavlov (2007), (8) Manzali et al. (2007), (9) Zharikov et al.(2007), (10) Koptsevich et al. (2001), (11) De Luca et al. (2005),(12) Martin et al. (1998), (13) Mignani et al. (2002), (14) Becker et al. (2006), (15) Zharikov et al. (2004), (16) Becker et al. (2004)
\end{table*}

We have  compared the spectrum  of the Vela  pulsar with those  of the
other  rotation-powered   INSs  for  which   either  medium-resolution
spectroscopy  or   multi-band  photometry  is   available  (Fig.   4).
Apparently,  the complexity  of the  spectral flux  distribution grows
with the INS  age.  For the young objects  the optical spectral energy
distribution is  dominated by a flat power-law  continuum which brings
the  signature  of  synchrotron  radiation  produced  by  relativistic
charged particles in the  neutron star's magnetosphere.  For the older
ones, a  Rayleigh-Jeans component, ascribed to  thermal radiation from
the cooling neutron star's surface, is also present.  In all cases, no
statistically significant evidence  of emission or absorption features
is found.  For all the objects  in Fig.4, Tab.  2 reports the measured
spectral index  $\alpha_O$ of the power-law  component, either derived
from spectroscopy or from broad-band photometry.  In those cases where
both values are available the  agreement is rather good, with the only
exception of PSR B0540--69.   However, as discussed in Serafimovich et
al. (2004),  its spectrum is unrecoverably polluted  by the background
of   the  surrounding,   bright  compact   ($\sim   4''$)  synchrotron
nebula. The comparison  of the optical spectral indeces  shows that it
is  difficult  to find  clear  spectral  templates  for different  INS
groups\footnote{We warn  here that the values of  $\alpha_O$ have been
computed  over slightly  different  wavelength ranges,  which makes  a
direct comparison more uncertain}.  This is shown in Fig.  5 (top left
panel), where we have plotted the optical spectral index $\alpha_O$ as
a function of the INS  spin-down age.  When available, we have assumed
as a reference the spectral index obtained from spectroscopy, with the
obvious exception of PSR B0540--69 (see above).

Although the  spectral index value  seems generally to  correlate with
the spin-down age, with  $\alpha_O \approx (0.12 \pm 0.04) Log(\tau)$,
the errors are such that most of the points are also consistent with a
costant distribution  around the  average value $<\alpha_O>  =0.44 \pm
0.4$.   Thus,  there is  no  firm evidence  for  an  evolution of  the
pulsars'  optical  spectral index  over  four  age  decades.  This  is
similar to what is found in  the X-rays, where the available data also
suggest  that the  value of  the  spectral index  $\alpha_X$ does  not
depend on the spin-down age  (see, e.g.  Becker \& Tr\"umper 1997).  A
possible anti  correlation between the optical spectral  index and the
rotational  energy loss $\dot  E$, with  $\alpha_O \approx  (-0.07 \pm
0.02) Log(\dot E)$, can  also be recognized in the  data (Fig.  5, top
right panel).   However, as before, the  large errors do  not allow to
draw any firm conclusion.  Finally, no trend can be recognized between
the optical spectral  index and the magnetic field  $B$ (Fig.5, bottom
left panel).

\begin{figure*}
\centering   
\includegraphics[bb=20 190 460 610 ,width=8.0cm,clip]{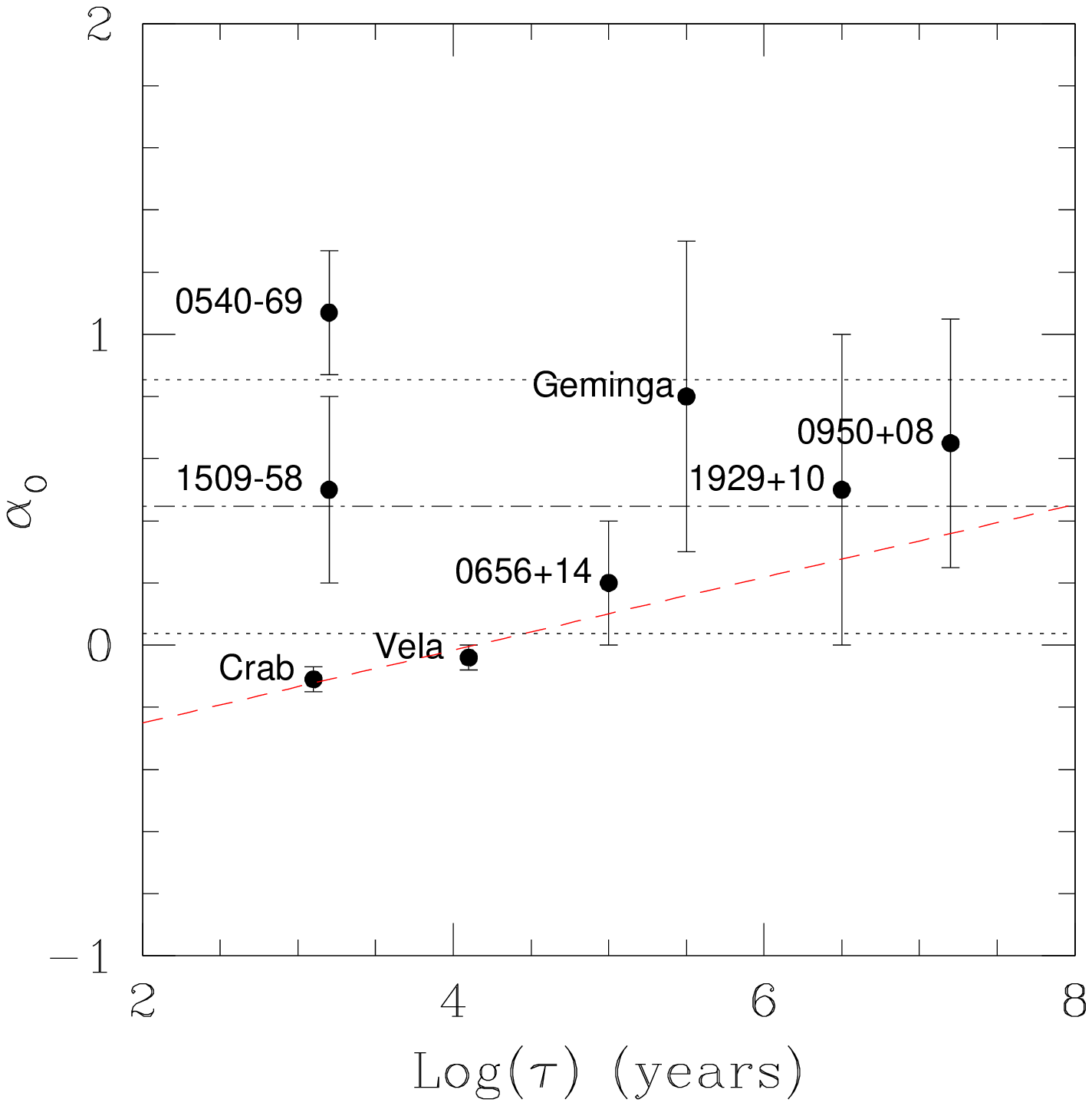}
\includegraphics[bb=20 190 460 610 ,width=8.0cm,clip]{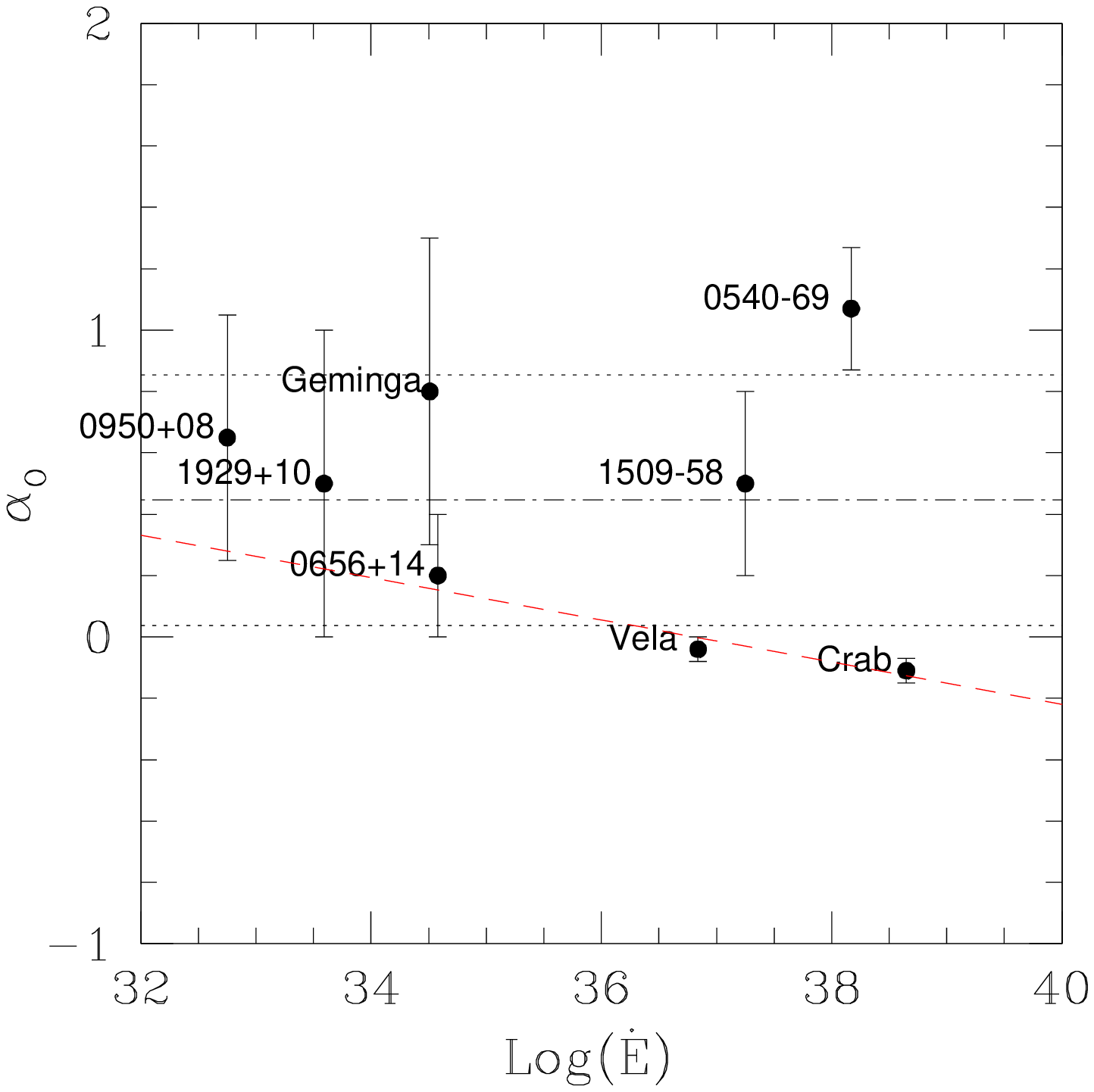}
\includegraphics[bb=20 190 460 610 ,width=8.0cm,clip]{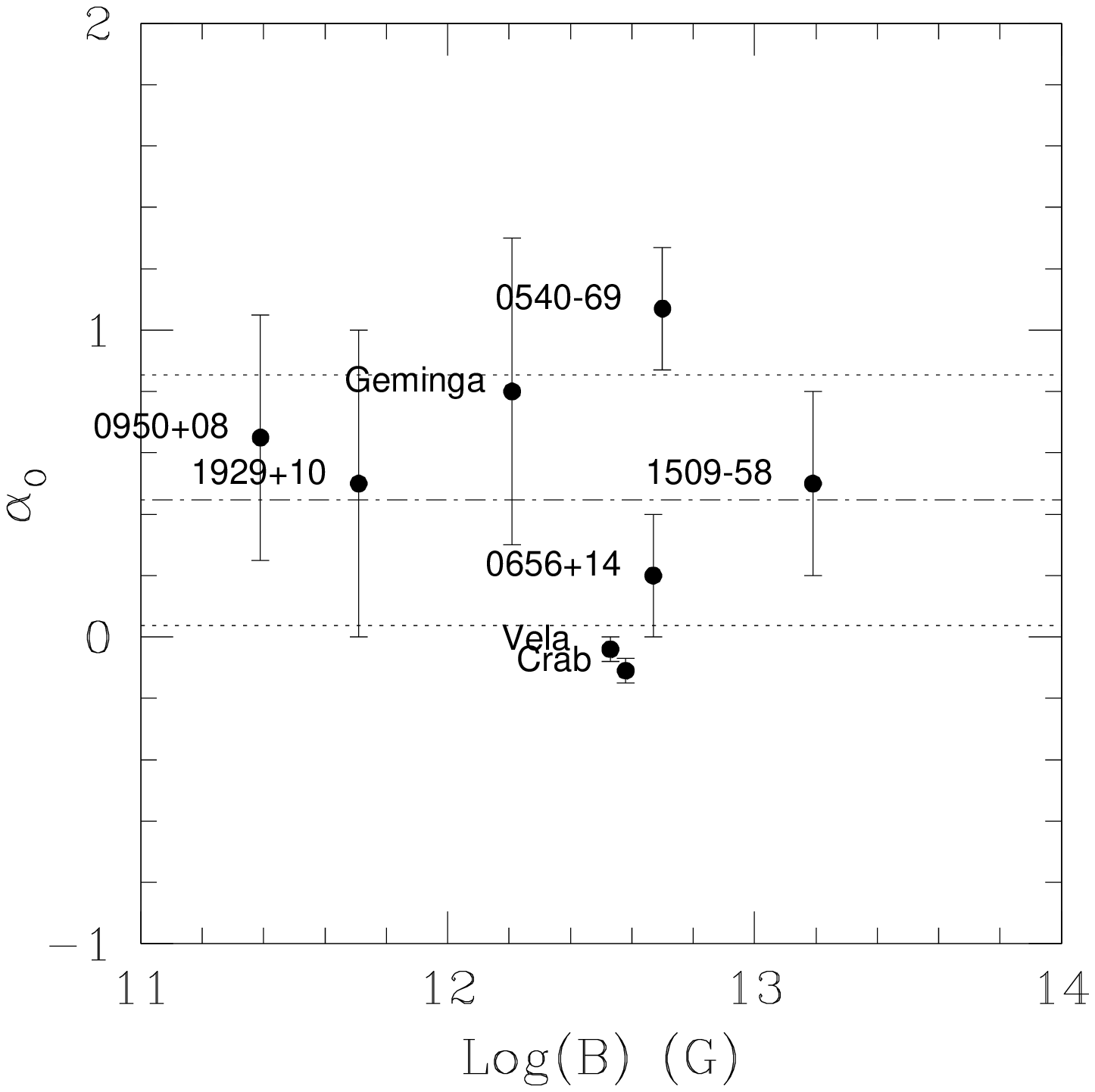}
\includegraphics[bb=20 190 460 610 ,width=8.0cm,clip]{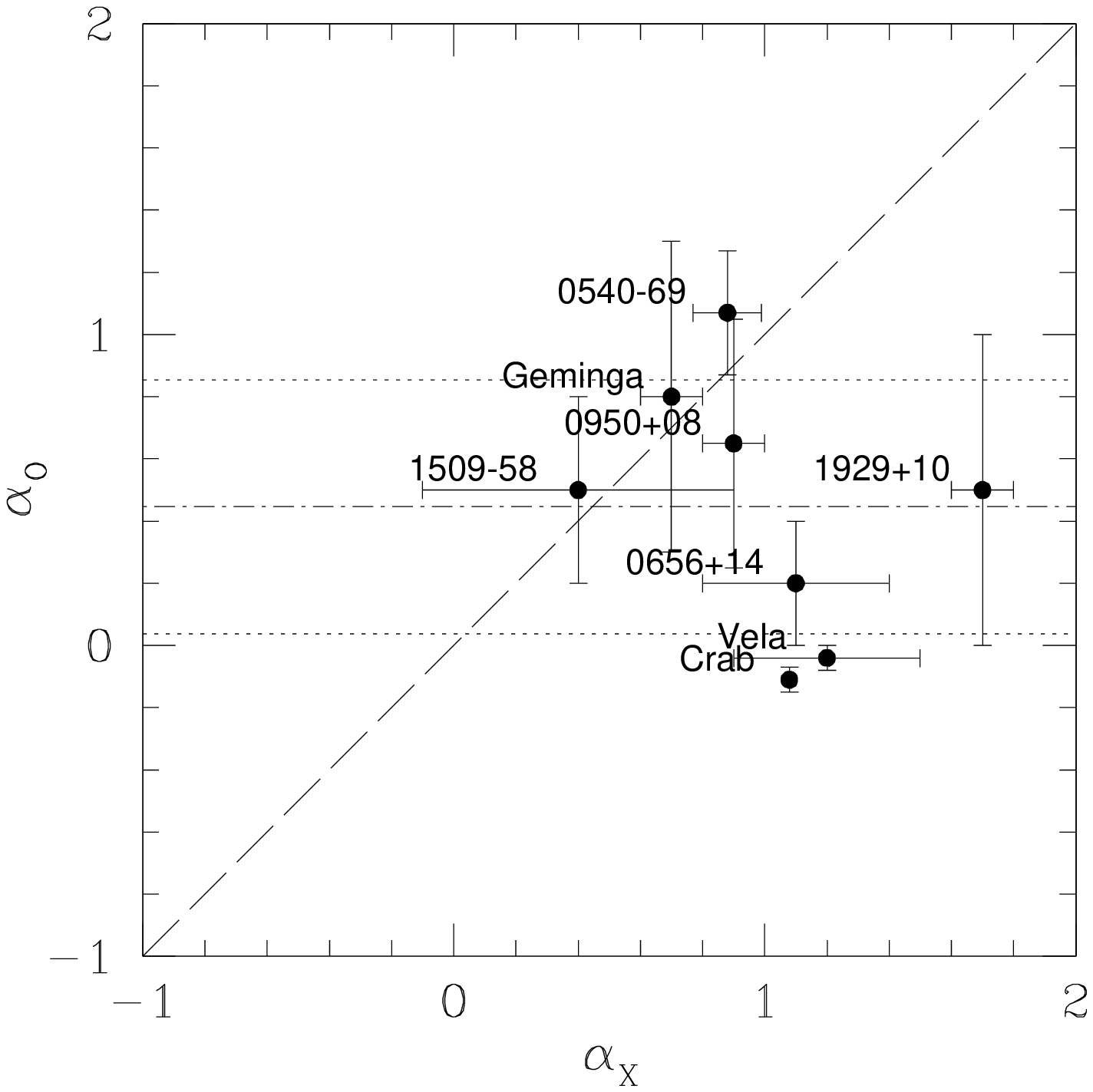}
\caption{Values of the optical spectral index $\alpha_O$ plotted as
a function  of the INS spin-down  age, rotational energy loss $\dot E$, the magnetic field $B$, and the X-ray spectral index $\alpha_X$ (top left to bottom right). The red dashed lines (top panels) represent
the linear fit to the points.  The dashed line (bottom right panel)
marks the case $\alpha_O = \alpha_X$. In all panels the dot-dashed and dotted lines correspond to the average value ($<\alpha_O> =0.44 \pm 0.4$) of the optical spectral index and to its $1 \sigma$ variation, respectively. }
\label{alpha_o}       
\end{figure*}

As it has been pointed out  (e.g. Mignani et al. 2004; Serafimovich et
al.   2004),   the  optical  and  X-ray   magnetospheric  emission  of
rotation-powered  pulsars  are  almost  never described  by  the  same
spectral parameters.  This  is clearly shown in Fig.   5 (bottom right
panel), where  we have plotted  the optical spectral  index $\alpha_O$
vs.  the  X-ray one $\alpha_X$.   As seen, the  optical power-law
index  is often less  steep than  the X-ray  one, suggesting  that the
spectra of rotation-power pulsars  undergo a general turnover at lower
energies.   In  particular, for the Crab the  slope of the power-law
swaps from  positive to  negative when passing  from the X-ray  to the
optical/IR domain,  underlying an even more  marked spectral turnover.
Furthermore,  no general  positive  or negative  correlation is  found
between $\alpha_O$ and $\alpha_X$,  which indicates that the X-ray and
optical magnetospheric emission, although  likely produced by the same
physical processes, are not  directly correlated.  This finding is
particularly  interesting in  comparison with  the  strong correlation
between the optical and X-ray luminosities of rotation-powered pulsars
noticed by Zharikov et al.  (2004) and by Zavlin and Pavlov (2004).

\section{Summary}

We have  presented the first optical spectroscopy  observations of the
Vela  pulsar.   The  pulsar's  spectrum  is characterized  by  a  flat
power-law  with spectral  index $\alpha  =-0.04 \pm  0.04$, consistent
with the values derived from broad-band photometry (Mignani \& Caraveo
2001;  Shibanov et  al.  2003;  Kargaltsev \&  Pavlov 2007).   We have
compared the newly  derived optical spectral index of  Vela with those
of all rotation-powered INSs for  which a power-law component has been
identified in the optical/IR spectrum. While a trend can be recognized
in the data, the large errors on the spectral index values for most of
the objects prevent  any claim for an evolution  of the magnetospheric
emission properties over four age  decades.  We also found no evidence
for  a correlation  between the  optical and  X-ray  spectral indeces,
which indicates  that the  X-ray and optical  magnetospheric emissions
are not directly correlated.  However,  we showed that in the majority
of cases,  the optical spectral index  is flatter than  the X-ray one,
suggesting a spectral turnover in the INSs' spectra at low energies.

\begin{acknowledgements}
RPM is supported by a PPARC Rolling Grant. SZ acknowledges  the support of the DGAPA/PAPIIT  project IN101506 and
of CONACYT 48493. We thank Werner Becker for the useful discussions and Oleg Kargaltsev for sending us the STIS/FUV pulsars fluxes. We thank our referee, Yury Shibanov, for his comments which helped to improve the quality of the manuscript. 
\end{acknowledgements}

\end{document}